\documentstyle[epsf,12pt]{article}
\begin{document}

\begin{titlepage}
\begin{flushright}
IFUP--TH/2009-21-r\\
\end{flushright}
~

\vskip .8truecm
\begin{center}
\Large\bf
On the semiclassical treatment of Hawking radiation
\end{center}

\vskip 1.6truecm
\begin{center}
{Pietro Menotti} \\
\vskip .8truecm
{\small\it Dipartimento di Fisica, Universit{\`a} di Pisa and\\
INFN, Sezione di Pisa, Largo B. Pontecorvo 3, I-56127}\\
\end{center}
                                                                               
\vskip 1.2cm

\centerline{November 2009}
                
\vskip 1.2truecm

\begin{abstract} 
In the context of the semiclassical treatment of Hawking radiation we prove
the universality of the reduced canonical momentum for the system of a massive
shell self gravitating in a spherical gravitational field within the
Painlev\'e family of gauges. We show that one can construct modes which are
regular on the horizon both by considering as hamiltonian the exterior
boundary term and by using as hamiltonian the interior boundary term. The late
time expansion is given in both approaches and their time Fourier expansion
computed to reproduce the self reaction correction to the Hawking spectrum.
\end{abstract}

\vskip 1truecm

\end{titlepage}

\section{Introduction}

In paper \cite{KW1} Kraus and Wilczek introduced a semiclassical treatment 
of Hawking radiation
by considering the mechanics of a thin self-gravitating shell of matter in a
spherical gravitational field. The interest of such an approach lies in the
fact that contrary to the usual external field treatment, energy conservation
is taken into account which allows to compute the self energy correction to
the Hawking formula. 

In \cite{KW1,KW2} the connection with Hawking radiation was obtained by
interpreting the exponential of the classical action as the modes of the
system. In this way in the first approximation the well known Hawking result
was re-obtained but in the full semiclassical approximation the self energy
correction were computed. In this analysis the key role is played by the
imaginary part of the canonical momentum which appears in the reduced
hamiltonian.

Subsequently \cite{PW} the results obtained in \cite{KW1,KW2} were given a new
interpretation as describing a tunneling phenomenon. Alternative formulae and
criticisms were proposed in this context
\cite{chowdhury,zerbini,akhmedov,akhmedova,zerbini1,zerbini2,zerbini3,
pizzi,belinski}. In the present paper however we shall go back to the
computation of the Bogoliubov coefficients as originally done in
\cite{KW1,KW2,KVK}.

In \cite{FLW,LWF} the problem of the dynamics of one thin shell of matter in
the framework of \cite{KW1,FMP} was examined critically and a precise
definition of the canonical momentum was given through a limit process
extending the treatment also to a massive shell of matter.

The procedure for deriving the reduced hamiltonian in \cite{KW1}, which is
obtained in implicit form, is usually considered as very complicated. 
In \cite{FM}
it was show that by introducing a proper generating function it is possible
to drastically simplify such a derivation and the procedure works 
also for massive shells. The method
allows also to extend the treatment to any finite number of massive or
massless shells which in their time development can cross. 
Moreover it was shown that the expression of the canonical momentum obtained in
\cite{KW1,FLW} holds in a more general setting and that no limit procedure is
necessary for obtaining it.

In this paper we shall examine two aspects of the problem: The first is the
universal character of the canonical momentum which appears in the reduced
hamiltonian within the class of the Painlev\'e gauges. This is done in
Sect.(\ref{thereducedactionsec}). Then we shall point out how the shell
dynamics can be obtained by considering as hamiltonian either the exterior
mass or an interior mass which appear in the boundary terms. This is done in
Sect.(\ref{equationssec}).

In Sect.(\ref{latetimessec}) we revisit the extraction of the Bogoliubov
coefficients from the semiclassical modes. A general treatment along this line
was given in the paper \cite{KVK}, but here we shall go back to the explicit
late-time development of the modes.  We show that
in constructing the semiclassical
modes which are regular at the horizon one can use either the exterior mass as
hamiltonian as was done originally in \cite{KW1,KVK} or an ``interior''
mass. The two hamiltonians are related to two different times, 
the exterior time,
which is the usual Killing time $t$ at space infinity, and the interior time
which we shall denote by $t'$.  Both approaches are consistent and they
produce the same result for the absolute value of the ratio of the $\beta$ to
the $\alpha$ Bogoliubov coefficients. In Sect.(\ref{conclusionssec}) we give a
discussion of the main conclusions.

In Appendix A we give some details of the calculations.

\section{The reduced action}\label{thereducedactionsec}

In this section we recall some features of the reduced action in the Painlev\'e
family of gauges. The spherically symmetric metric is written as
\begin{equation}\label{metric}
ds^2 = -N^2dt^2+L(dr+N^rdt)^2 + R^2 d\Omega^2
\end{equation}
where the functions $N, N^r, L, R$ can be consistently assumed to be continuous
functions of $r$ and $t$ \cite{FLW}.
In \cite{FM} the function
\begin{equation}\label{generalF}
F=R L \sqrt{\left(\frac{R'}{L}\right)^2 -1+ \frac{2{\cal M}}{R}} +
RR'\log\left(\frac{R'}{L}- \sqrt{\left(\frac{R'}{L}\right)^2 -1+ 
\frac{2{\cal M}}{R}}\right)
\end{equation}
was introduced which has the remarkable property of generating the conjugate 
momenta as 
follows
\begin{equation}
\pi_L =\frac{\delta F}{\delta L}=\frac{\partial F}{\partial L}
\end{equation}
\begin{equation}\label{piRfromF}
\pi_R = \frac{\delta F}{\delta R}=\frac{\partial F}{\partial
R}-\frac{\partial }{\partial r}\frac{\partial F}{\partial R'}. 
\end{equation}

The Painlev\'e family of gauges is identified by the choice
$L=1$ in the metric (\ref{metric}) where the generating function $F$ assumes 
the form
\begin{equation}\label{specialF}
F = R W(R,R',{\cal M}) +
RR'({\cal L}(R,R',{\cal M}) - {\cal B}(R,{\cal M}))
\end{equation}
where
\begin{equation}\label{WcalL}
W(R,R',{\cal M})= \sqrt{R'^2-1+\frac{2{\cal M}}{R}};~~
{\cal L}(R,R',{\cal M}) = \log(R'-W(R,R',{\cal M}))
\end{equation}
and
\begin{equation}\label{calB}
{\cal B}(R,{\cal M}) = \sqrt{\frac{2{\cal M}}{R}}+\log(1-\sqrt{\frac{2{\cal
M}}{R}}). 
\end{equation}
In going from eq.(\ref{generalF}) to eq.(\ref{specialF}) we exploited the
freedom of adding a total derivate to $F$ with the result that the function
(\ref{specialF}) has the useful property of vanishing wherever $R'=1$.

One is still left with the freedom to impose a gauge condition on $R$. In
\cite{FM} several choices were examined which can be characterized by a
deformation function, of bounded support around the shell position $\hat r$
\begin{equation}
R(r) = r+ c g(r-\hat r).
\end{equation}
We shall call a gauge of the outer type if $g(x)=0$ for $x\geq 0$; 
if $g(x)=0$ for
$x\leq 0$ the gauge is called of inner type. 
There are however other choices in 
which $g$ is not vanishing in a neighborhood of $0$ both for positive and 
negative argument.

It is useful to start from the general form of the action on a bounded region
of space-time as given in \cite{HH} to which the shell action 
\begin{equation}\label{shellaction}
S_{shell}=\int_{t_i}^{t_f} dt ~\hat p~\dot{\hat r}
\end{equation}
is added and using the generating function $F$, one
can derive in a straightforward way \cite{FM} the reduced action, boundary
terms included
\begin{equation}\label{reducedaction}
\int_{t_i}^{t_f} \left(p_{c}~ \dot{\hat r} -\dot M(t)\int_{r_0}^{\hat
r(t)}\frac{\partial F}{\partial M} dr+ \left.(-N^r \pi_L +
NRR')\right|^{r_m}_{r_0}\right)dt 
\end{equation}
where the outer gauge $g(x)=0$ for $x \geq 0$ has been adopted in the 
Painlev\'e family of gauges.  
The boundary terms are just the one given in the paper \cite{HH} computed for
the spherically symmetric problem at hand. They are equivalent to
\begin{equation}
-H N(r_m) + M N(r_0)
\end{equation} 
where as a consequence of the constraints $M$ and $H$  are constant in $r$
except at the position of the shell $\hat r$. 
$M$ is the interior mass while $H$ denotes the exterior mass.
Furthermore as a consequence of the gravitational equations combined with the
constraints, $M$ and $H$ are also constant in $t$. 

However this does not entail one to drop the $\dot M$
term appearing in eq.(\ref{reducedaction}), because such a time constancy can 
be used only after having derived the equations of motion. 
In deriving (\ref{reducedaction}) the so called outer gauge has been used
i.e. in which the radial function $R(r)$ appearing in the metric
(\ref{metric})  is equal to
$r$ for $r>\hat r$, and is deformed below $\hat r$ on a finite tract with a
smooth function $g$ with $g(x)=0$ for $x\geq 0$, $g'(0-)=1$ and
\begin{equation}\label{Rfunction}
R(r,t) = r+\frac{V(t)}{\hat r} g(r-\hat r(t))
\end{equation}
due  to the fact that as a consequence of the constraints
\begin{equation}\label{DeltaR1}
\Delta R'\equiv R'(\hat r+0)-R'(\hat r-0) =
-\frac{V}{R};~~~~V=\sqrt{{\hat p}^2+m^2}
\end{equation}
being $m$ the mass of the shell.
The imposition of the constraints is the reason why one cannot adopt 
$R=r$ for all $r$.

In the inner gauge a term of type
$\dot H$ will appear while in a generic gauge both $\dot M$ and $\dot H$ terms
will appear. Moreover the constraints impose
\begin{equation}\label{DeltapiL}
\Delta \pi_L=-\hat p
\end{equation}
being
\begin{equation}
\pi_L = R \sqrt{(R')^2 -1 + \frac{2{\cal M}}{R}}\equiv R
W(R,R',{\cal M})
\end{equation}
where by ${\cal M}$ we denote the mass which is function constant in $r$ for
$r>\hat r$ and also for $r<\hat r$ but discontinuous at $\hat r$.

The general form of $p_c$ is given by \cite{FM}
\begin{equation}
p_c = \hat r(\Delta{\cal L}-\Delta{\cal B})
\end{equation}
with ${\cal L}$ and ${\cal B}$ given by eqs.(\ref{WcalL},\ref{calB}).
Using relations (\ref{DeltaR1},\ref{DeltapiL}) $p_c$ becomes 
\begin{equation}\label{pcouter}
p_c(\hat r)= \sqrt{2M\,\hat r}-\sqrt{2H \,\hat r}-\hat
r\log\left(\frac{\hat r+\sqrt{{\hat p}^2+m^2}-\hat p-
\sqrt{2H \,\hat r}}{\hat r-\sqrt{2M \,\hat r}}\right)
\end{equation}
where \cite{FLW,FM} $\hat p$ is given implicitly by the
solution of the equation
\begin{equation}\label{fundamentalH}
H-M= V +\frac{m^2}{2\hat r}-\hat p\sqrt{\frac{2H}{\hat r}}.
\end{equation}
We stress that no limit procedure is necessary to obtain 
(\ref{pcouter},\ref{fundamentalH}) which
hold with any deformation to the left of $\hat r$.

In the inner gauges in 
which the deformation is taken to the right of $\hat r$ one
obtains for the canonical momentum $p_c$
\begin{equation}
p_c^i= \sqrt{2M\,\hat r}-\sqrt{2H \,\hat r}-\hat
r\log\left(\frac{\hat r-\sqrt{2H \,\hat r}}
{\hat r- V^i + \hat p^i-\sqrt{2M\hat r}}\right)
\end{equation}
where now $\hat p^i$ is given by the implicit equation
\begin{equation}\label{fundamental}
H-M=V^i-\frac{m^2}{2\hat r}-\hat p^i \sqrt{\frac{2M}{\hat r}}
\end{equation}
which is different from eq.(\ref{fundamentalH}). More general gauges can also
be considered in which the discontinuity (\ref{DeltaR1}) in $R'$ at $r=\hat r$
is 
partly due to a deformation on the right and to  a deformation on the left of
$\hat r$. 
It is possible to show 
that actually $p_c$ is independent from any gauge choices in
which $L=1$. In fact after some algebra given in Appendix A, we find
\begin{equation}\label{pcmassive}
p_c = \sqrt{2 M\hat r}-\sqrt{2 H\hat r }
- \hat r \log\frac{\hat r -\sqrt{2H \hat r}}{\hat r -\sqrt{2M \hat r}}
-\hat r\log\frac{\hat r - H - M -\frac{m^2}{2 \hat r} 
-\hat r \sqrt{C}}{\hat r -2H}
\end{equation}
being $C$ the discriminant
\begin{equation}
C =(\frac{H-M}{\hat r})^2+\frac{m^2}{\hat r^2}(\frac{H+M}{\hat r}-1)+
\frac{m^4}{4\hat r^4}.
\end{equation}
Notice that the argument of the second logarithm becomes $1$ for $m^2=0$ and
in this situation the last term in (\ref{pcmassive}) disappears.

The mass $m$ of the shell does
not play a really important role in the phenomena we shall examine in the
following, the reason being that, as seen from eq.(\ref{explicitphat}) of the
Appendix at the horizon $r=2 H$ where $p_c$ is singular $\hat p$ diverges 
and it cancels
the term $V$ and we know that Hawking radiation which is a late time
phenomenon depends on the behavior of the modes at the horizon. Moreover
eq.(\ref{Imofintegral}) below holds also for $m\neq 0$.

For $m=0$
\begin{equation}\label{pcmassless} 
p_c= \sqrt{2M\,\hat
r}-\sqrt{2H \,\hat r}-\hat r\log\left(\frac{\hat r- \sqrt{2H \,\hat r}}{\hat
r-\sqrt{2M \,\hat r}}\right) 
\end{equation} 
which is the original result by
Kraus and Wilczek \cite{KW1}.  The space-part of the semiclassical mode is 
given by 
\begin{equation}\label{spacepart} 
u(r) ={\rm const}~
\exp(i(\int^{\hat r} p_c(\hat r) d\hat r + {\rm const})) 
\end{equation}
where the additive constant can be taken also to depend on $H,~M$ and
contributes only to a phase factor in the semiclassical mode. 
For $m=0$ the integral
appearing in eq.(\ref{spacepart}) can be easily computed.  One finds 
\begin{equation}\label{pczeromass}
\int p_c(\hat r) d\hat r = f(\hat r,M)-f(\hat r,H)
\end{equation} 
where 
\begin{equation} 
f(\hat r,M) =\frac{\hat
r^2-4M^2}{2}\log(\sqrt{\hat r} - \sqrt{2M}) +\frac{\hat r-2M}{2} (\sqrt{2M\hat
r}+\frac{\hat r}{2}). 
\end{equation} 
The large $\hat r$ behavior of $p_c$ is
\begin{equation} 
\lim_{\hat r \rightarrow \infty }p_c(\hat r) = H-M=\sqrt{\hat p^2(+\infty)+
m^2}\equiv \omega.
\end{equation}
The $p_c$ as a function of $\hat r$ develops an imaginary part $i\pi\hat r$ in
the ``gap'' $2M,2H$. This is evident from eq.(\ref{pcmassless}) 
but in \cite{FM} it was
shown to be true also in presence of mass i.e. for eq.(\ref{pcmassive}) 
independently
of the value of $m<H-M$; the last is the necessary and sufficient condition
for a massive shell to reach space infinity. One has as a consequence
\begin{equation}\label{Imofintegral} 
{\rm Im} \int p_c d\hat r = 2 \pi(H^2-M^2). 
\end{equation} 
Thus we have shown that $p_c$ is independent of the gauge choice, within the
family of the Painlev\'e gauges, and that eq.(\ref{Imofintegral}) holds for
also for $m\neq 0$.  

The exponential of minus Eq.(\ref{Imofintegral}) was given the interpretation
of a tunneling amplitude by Parikh and Wilczek \cite{PW}. Criticism and
alternative proposal for the tunneling amplitudes were given in the literature
\cite{chowdhury,zerbini,akhmedov,akhmedova,zerbini1,zerbini2,zerbini3,pizzi,
belinski}.

Here however we shall not pursue this line of thought and instead still
employing eqs. (\ref{pcouter},\ref{fundamentalH},\ref{pczeromass}) we shall go
back to the calculation of coefficients of the Bogoliubov transformation.

The expansion in $\omega = H -M$ of eq.(\ref{pczeromass}) to second
order in $\omega$ is
\begin{eqnarray}\label{omega2expansion}
\omega\left(M + 2 \sqrt{2 M \hat r}+ \hat r + 4 M \log(\frac{\sqrt{\hat r}
-\sqrt{2 M}}{\sqrt{2 M}})\right)
+\\
\frac{\omega^2}{2}
\left(1 - \frac{2\sqrt{2M}}{\sqrt{\hat r}-\sqrt{2M}}+\frac{\sqrt{2\hat r}}
{\sqrt{M}}+
4 \log(\frac{\sqrt{\hat r}-\sqrt{2 M}}{\sqrt{2M}})\right)+ \cdots
\end{eqnarray}
What matters in the late time emission is the behavior of the mode at the
horizon which is given by
\begin{equation}
4M \omega\log(\frac{\hat r - 2M}{4M})+4M \omega^2\left( \frac{1}{2M }
\log(\frac{\hat r -2M}{4M})-
\frac{1}{\hat r - 2M}\right)+\cdots
\end{equation}
We adopt the usual notation \cite{KW1} for the expansion of the scalar field
$\phi$ 
\begin{equation}
\phi =\int \frac{d\omega}{\sqrt{2\omega}}(u_\omega(\hat r)e^{-i\omega
t}a(\omega)+ u^*_\omega(\hat r)e^{i\omega t}a^+(\omega)) 
\end{equation}
where now $t$ is the Painlev\'e time and
$ u_\omega(\hat r)$ is given by eq.(\ref{spacepart}). 
The same field $\phi$ can be expanded in modes which are regular at the
horizon \begin{equation} 
\phi =\int \frac{dk}{\sqrt{2 k}}(v_{k}(\hat r,t) b(k)+
v^*_{k}(\hat r,t) b^+(k)). 
\end{equation} 
The behavior of such $v$ modes at
the horizon is given by translating the expression \cite{unruh} 
\begin{equation}
v_{k}= \frac{1}{\sqrt{2\pi}} e^{ik(X-T)} 
\end{equation} 
with
\begin{equation} X=\frac{M}{2}(V-U);~~~~T=\frac{M}{2}(V+U);~~~~ 
U=-e^{-\frac{u}{4M}};~~~~V=e^{\frac{v}{4M}};~~~~
\end{equation} 
into the Painlev\'e coordinates, being
\begin{equation} u=t_s-\hat r_*;~~~~v=t_s+\hat r_*;~~~~ \hat r_*=\hat
r+2M\log(\frac{\hat r}{2M}-1)  
\end{equation}
and $t_s$ the Schwarzschild time 
\begin{equation} 
t_s = t-2\sqrt{2M\hat r}-2M\log\frac{\sqrt{\hat r}-\sqrt{2M}}
{\sqrt{\hat r}+\sqrt{2M}}+{\rm const}.
\end{equation} 
Near the horizon one finds
\begin{equation}\label{regularmode}
 v_{k}(\hat r,t) =\frac{1}{\sqrt{2\pi}}
\exp(ik(\hat r-2M)e^{-\frac{t}{4M}}).
\end{equation}
It is essential in obtaining eq.(\ref{regularmode}) to adopt the 
Painlev\'e time. In fact
such regular modes will be subsequently analyzed in terms of the $u$-mode
which live also on the Painlev\'e background.
 
The usual technique for extracting the Bogoliubov coefficient \cite{BD} is
that of projection of the regular $v$-modes on the $u$-modes by space
integration. Here the main difference w.r.t. the usual treatment is that the
background metric is now the Painlev\'e metric with the mass $M$. Despite
being the Painlev\'e metric a non static metric still it describes a static
space time and thus due to the invariance of the space integrals the usual
formalism for computing the scalar products works and there is no need to
adopt the general form for stationary spaces \cite{AM}. Such a scalar product
is given by
\begin{equation}\label{genscalarproduct}
-i\int(\psi_2^*\partial_\rho \psi_1 - \psi_1\partial_\rho \psi_2^*)g^{\rho t}
\sqrt{-g}~\varepsilon_{tr\theta\phi} ~d\hat r d\theta d\phi 
\end{equation}
where the integration region is outside the horizon. 
Using $\sqrt{-g}={\hat r}^2 \sin\theta$ we have for eq.(\ref{genscalarproduct})
\begin{equation}
4\pi i\int_{2M}^\infty(\psi_2^*\partial_t\psi_1-\psi_1\partial_t \psi_2^*)
{\hat r}^2 d\hat r  
-4\pi i\int_{2M}^\infty(\psi_2^*\partial_r\psi_1-\psi_1\partial_r\psi_2^*)
N^r {\hat r}^2 d\hat r;~~N^r=g^{rt}=\sqrt{\frac{2M}{\hat r}}.
\end{equation} 
Taking into account that eq.(\ref{spacepart},\ref{regularmode}) are reduced 
radial modes we find for the  most singular term in the integrand giving 
the scalar
product of $v_k$ with $u_\omega$
\begin{equation}
\frac{\sqrt{M}}{\pi} \int_0^\infty e^{-ikx\tau}e^{4i\omega M \log x-i\omega t}
\frac{dx}{x} 
\end{equation}
where such term originates from $g^{rt} v_k^*\partial_r u_\omega$ and 
$\tau = e^{-\frac{t}{4M}}$, $x=\hat r-4M$ and we took into account the
normalization of the $u_\omega$.

Integrating in $x = \hat r-2M$ we have
\begin{equation}
\beta^*_{k\omega} = \frac{\sqrt{M}}{\pi} \sqrt{\frac{\omega}{k}}\int_0^\infty 
e^{ikx\tau}
e^{4iM\omega \log x -i\omega 
t} \frac{dx}{x}=  \frac{\sqrt{M}}{\pi}  \sqrt{\frac{\omega}{k}} 
\int_0^\infty e^{ikx\tau} e^{4iM\omega \log (x\tau)} 
\frac{d(x\tau)}{x\tau}
\end{equation}
while $\alpha_{k\omega}$ is obtained changing $\omega$ in $-\omega$.
On the other hand one can extract the Bogoliubov coefficients also performing
a time Fourier transform of the regular modes at fixed $r$
\begin{equation}
\beta^*_{k\omega} = -\frac{1}{4\sqrt{M}\pi}\sqrt{\frac{\omega}{k}}
\int_{-\infty}^\infty e^{ikx\tau}
e^{4iM\omega \log x -i\omega 
t} dt = \frac{\sqrt{M}}{\pi} \sqrt{\frac{\omega}{k}}\int_0^\infty 
e^{ikx\tau} e^{4iM\omega \log (x\tau)} \frac{d(x\tau)}{x\tau}
\end{equation}
which also shows that the result does not depend on the value of $\hat r$. The
independence of the time Fourier transform from $\hat r$ can be proved on
general grounds for the exact modes.
Thus we have that the first order term reproduces the well-known Hawking
integrals and the treatment outlined above is obviously equivalent to the
original external field treatment even if here it is developed in the
Painlev\'e reference frame. 

There are several reasons why one cannot exploit the second order term in the
expansion (\ref{omega2expansion}). The second order contains a non integrable
singularity in $\hat r$; in addition it would not be proper to compute the
scalar 
product of eq.(\ref{omega2expansion}) with the regular mode $v$
eq.(\ref{regularmode}) as also here one should consider the correction due to
the back reaction. In addition one cannot employ any longer the Painlev\`e
background metric characterized by $M$ but one should also consider the
corrections to it due to the presence of the shell.
For this reasons in Sect.(\ref{latetimessec}) we shall  adopt the time Fourier
transform technique as done in \cite{KW1}

\section{The equations of motion}\label{equationssec}

We recall that in deriving the equations of motion one can vary $H$ keeping
$M$ as a fixed datum or vary $M$ keeping $H$ as a fixed datum \cite{FM}.  The
first procedure is much simpler if one adopts the outer gauge because the term
in $\dot M$ is zero in eq.(\ref{reducedaction}) and one obtains
\begin{equation}\label{eqmotion1}
\dot{\hat r} \frac{\partial p_c}{\partial H}-N(r_m)=0
\end{equation}
where $N(r_m)$ does not depend on $r_m$ for $r_m>\hat r$ and the usual
normalization $N(r_m)=1$ corresponds to identifying $t$ with the Killing time 
at space infinity. In this case eq.(\ref{eqmotion1}) becomes \cite{FLW,FM}
\begin{equation}\label{exteriorequation}
\frac{d\hat r}{dt} = \left(\frac{\hat p}{V}-\sqrt{\frac{2H}{\hat r}}\right).
\end{equation}

The procedure in which one varies $M$ keeping $H$ fixed is more complicated
due to the presence of the term in $\dot M$ in action (\ref{reducedaction}) 
and which cannot be neglected to obtain the correct equations of motion. 
It was explicitely proven in \cite{FM} that one gets 
the same equations of
motion (\ref{exteriorequation}) if one maintains the normalization
$N(r_m)=1$. On the other hand if one adopts the normalization $N(r_0)=1$ one
obtains the equations of motion w.r.t. the time which flows at $r_0$ and which
we shall denote by $t'$. The calculation is most easily performed in the inner
gauge obtaining 
\begin{equation}
\frac{d\hat r}{dt'} \frac{\partial p_c}{\partial M}+N(r_0)=0;~~~~{\rm with}~~~
\frac{\partial p_c}{\partial M}=
\left(\frac{\hat p^i}{V^i}-\sqrt{\frac{2M}{\hat r}}\right)^{-1} 
\end{equation}
and thus
\begin{equation}\label{interiorequation} 
\frac{d\hat r}{dt'} =
\left(\frac{\hat p^i}{V^i}-\sqrt{\frac{2M}{\hat r}}\right). 
\end{equation}
Thus we see that, due to the sign of the boundary term in
eq.(\ref{reducedaction}), choosing $N(r_0)=1$ gives $-M$ as the hamiltonian in
this scheme.  

We shall use both calculational schemes in the next section.  The values of
$H$ and $M$ which on shell are constant of motions give the values of the
mass contained in $r<r_m$ and the mass contained in $r<r_0$. 
Despite being the coordinate $r$ one dimensional, $r_0$ and $r_m$ do
not play an exactly symmetrical role as in deriving the action
the jacobian $R^2$ plays an essential role.

\section{The late time expansion}\label{latetimessec}

In this section we give a simplified derivation of the late time expansion of
the outgoing modes which are regular at the horizon. In constructing such
regular modes one can use the scheme in which $H$ plays the role of the
hamiltonian while $M$ is a given fixed parameter.  This is the scheme followed
in \cite{KW1,KVK}. One can also construct regular modes by giving $H$ as a
fixed parameter and using $M$ as hamiltonian as was described in Sect.
(\ref{equationssec}) in the second derivation of the equations of motion.

We start form the case in which $M$ is a given constant, while $H$ plays the
role of the hamiltonian.

The distinguishing feature of Hawking radiation is that of being a late time
effect. Below we shall compute with a simple technique the late time expansion
of the regular mode confining ourselves to the $(\omega/M)^2$ corrections to
the Hawking result. To that end it is sufficient to compute just the first
order correction in the late time expansion. In the expansion the terms are
classified in power of $\tau=e^{-\frac{t}{4M}}$ and powers of $t$. We shall
need only the $O(\tau)$ and $O(t\tau^2)$ terms which are very simply
extracted.

By keeping only the singular terms in $p_c$ and in the time development we
have the equations which were used in \cite{KW1} 
\begin{equation}\label{pcsystem} 
p_c =
-\hat r\log{\frac{\sqrt{\hat r}-\sqrt{2H}}{\sqrt{\hat r}-\sqrt{2M}}} 
\end{equation}
\begin{equation}\label{schwtime} 
t = 4H\log(\sqrt{\hat r}-\sqrt{2H})-4H\log(\sqrt{\hat r(0)}-\sqrt{2H}).
\end{equation} 
Furthermore regularity is obtained by imposing that at $t=0$
$p_c=k$ with $k>0$.  Thus we have the further restriction 
\begin{equation}\label{kcondition}
0<k
= - \hat r(0)\log\frac{\sqrt{\hat r(0)}-\sqrt{2H}}{\sqrt{\hat r(0)}-\sqrt{2M}}
\end{equation} 

The action is given by \cite{KW1,KVK}
\begin{equation}\label{regularmode2} 
k\hat r(0)+\int_0^t (p_c(\hat r(t'),H(t),M,k)\dot{\hat r}(t') - H(t))dt'
\end{equation} 
where for once we wrote explicitely the dependence of $p_c$ on time. $p_c$ is
computed on the solution of the equation of motion with the conditions $p_c=k$
at $t'=0$ and $\hat r(t)=\hat r$, being $\hat r$ a fixed value of the
shell coordinate outside the horizon and $t$ an arbitrary time. In the
procedure we 
shall develop below the only thing we shall need is the function $H(t)$ which
we shall compute in the late time expansion.
In the previous as in the following equations down to eq.(\ref{blackbody})
$H$ stays for $H(t)$.

From eq.(\ref{schwtime}) we obtain
\begin{equation}\label{r0}
\sqrt{\hat r(0)}-\sqrt{2H}=(\sqrt{\hat r}-\sqrt{2H})e^{-\frac{t}{4H}} 
\end{equation}
which substituted in eq.(\ref{kcondition}) gives the implicit equation
for the time evolution of $H(t)$ 
\begin{equation}\label{Hequation}
k=(\sqrt{2H}+(\sqrt{\hat r}-\sqrt{2H})e^{-\frac{t}{4H}})^2
\log\frac{(\sqrt{\hat r}-\sqrt{2H})
e^{-\frac{t}{4H}}+\delta_H}{(\sqrt{\hat r}-\sqrt{2H})e^{-\frac{t}{4H}}}
\end{equation} 
with
\begin{equation}
\delta_H = \sqrt{2H}-\sqrt{2M}.
\end{equation}

From eq.(\ref{Hequation}) we see that $t$ going to $+\infty$, $\sqrt{2H}$ goes
over to the finite value $\sqrt{2M}$ because $\delta_H$ has to go to zero.

To lowest order we have
\begin{equation}
\delta_H= c_H  T
\end{equation}
with $T\equiv e^{-\frac{t}{4H}}$ which
substituted in eq.(\ref{Hequation}) gives for $c_H$
\begin{equation}
c_H= (e^{\frac{k}{2M}}-1)L;~~~~L\equiv \sqrt{\hat r}-\sqrt{2M}. 
\end{equation}
Having determined the constant $c_H$ we can go over to the next
term. By simply expanding we obtain for large $t$
\begin{equation}\label{mainexpansion}
\frac{t}{4H(t)} = \frac{t}{4(M+\sqrt{2M} c_H e^{-\frac{t}{4M}})}=
\frac{t}{4M}\left(1-\frac{\sqrt{2M}}{M}c_H \tau+O(\tau^2)\right) 
\end{equation} 
giving
\begin{equation}
\delta_H \equiv \sqrt{2H}-\sqrt{2M} = c_H \tau+(c_H
\tau)^2\frac{t}{(2M)^{3/2}}+O(\tau^2) 
\end{equation}
or
\begin{equation}\label{Hoft}
H(t)=M+\sqrt{2M}c_H\tau +\frac{t}{2M}(c_H\tau)^2+O(\tau^2)
\end{equation}
where $\tau = e^{-\frac{t}{4M}}$.

The last obtained relation (\ref{Hoft}) is what is necessary and sufficient to
compute the  
$\omega^2$ corrections in the late time framework. In fact we are
interested only in the time dependence of the mode given  by the exponential
of $i$ 
times the expression (\ref{regularmode2}) and we have \cite{KVK} 
\begin{equation}
\frac{\partial S}{\partial t} = -H(t) 
\end{equation}
which holds also with the boundary conditions (\ref{kcondition}) as can be
explicitely verified. 
Thus the time development of $S$ is given by
\begin{equation}
S= f(\hat r) -\int^t H(t') dt'.
\end{equation}
Integrating we find
\begin{equation}
\int^t H(t')dt' = {\rm const}+Mt-4M \sqrt{2M}\tau_1-t\tau_1^2+..
\end{equation}
where for notational simplicity we set $\tau_1 = c_H \tau$.

Thus $S$ at large times behave as $-Mt$ independently of $k$. 
On the other hand the Fourier time
analysis of $e^{iS}$ contains frequencies which are above and below such value
$M$ and this is the well known fact that the mode of the system which is
regular at the horizon does not represent an eigenvalue of the energy as
measured by a stationary observer at space infinity.  The deviations from the
value $M$ represent the positive and negative frequency content of the
radiation mode. Thus after subtracting the background frequency $M$
we have in the time Fourier analysis the exponent
\begin{equation}\label{exponentH}
i(S -M t\pm\omega t) = i(f(r) - \int^t H(t') dt'+ Mt  \pm \omega t). 
\end{equation}
The saddle point $-H(t)+M \pm \omega=0$ is given by
\begin{equation}\label{saddlepoint}
\sqrt{2M}\tau_1+\frac{t}{2M}\tau_1^2 \mp \omega=0.
\end{equation}
The upper sign refers to the calculation of 
the $\alpha$ coefficients while the lower sign refers to the $\beta$
coefficients. For the upper sign we see that the saddle point is real thus
giving in the saddle approximation the contribution 1 to the modulus of 
$\alpha$. For the
lower sign i.e. for the $\beta$ coefficient, the exponent (\ref{exponentH}) 
becomes, using eq.(\ref{saddlepoint})
\begin{equation}\label{saddlevalue}
i[4M\sqrt{2M}\tau_1+t(\tau_1^2-\omega)]=-i[4M \omega +
t\omega(1+\frac{\omega}{2M})].
\end{equation}
Thus to compute $|\beta_{k\omega}/\alpha_{k\omega}|$ to second order in
$\omega$ we need ${\rm Im}t$ to first order in $\omega$.

We have
\begin{equation}
\sqrt{2M}\tau_1=-\omega(1+\frac{t}{4M^2}\omega)= -\omega(1-\frac{\omega}{M}
\log(-\frac{\omega}{\sqrt{2M}c_H}))
\end{equation}
so that to first order in $\omega$ we have
\begin{equation}
t= -4M (1-\frac{\omega}{M})\log(-\frac{\omega}{\sqrt{2M}c_H})
\end{equation}
giving 
\begin{equation}
{\rm Im}t = -4\pi M(1-\frac{\omega}{M})
\end{equation}
Combining with eq.(\ref{saddlevalue})
\begin{equation}\label{betaoveralphaH}
\left|\frac{\beta_{k\omega}}{\alpha_{k\omega}}\right| = 
e^{-4\pi M\omega(1-\frac{\omega}{2M})}.
\end{equation}
This is the explicit derivation through the late time expansion of the result
obtained in \cite{KVK} which corrects a previous result in \cite{KW1}. From
eq.(\ref{betaoveralphaH}) through a standard procedure \cite{KW1} which
exploits the wronskian property of the Bogoliubov coefficients one obtains the
spectrum of the radiation
\begin{equation}\label{blackbody}
F(\omega)d\omega=\frac{d\omega}{2\pi}\frac{1}{e^{8\pi
M\omega (1-\frac{\omega}{2M})}-1}.
\end{equation}

In the above treatment the parameter $M$ which characterizes the regular
modes has to be identified with the semiclassical mass of the system which
emits the radiation because it is the value of the hamiltonian $H(t)$ at large
values of the time $t$ and $H(t)$ is the boundary term at $r_m$ which is
related to the total energy of the system i.e. the exterior mass.

\bigskip

In order to clarify the issue we want now to repeat the analysis from a
different view point, i.e. by considering $H$ as a given parameter and $M$,
i.e. the interior mass, as hamiltonian.  Actually from the boundary term in
eq.(\ref{reducedaction}) with $N(r_0)=1$ we see that the hamiltonian i.e. the
generator of the time translations is $-M$ as we discussed at the end of
Sect.(\ref{equationssec}). 
Thus from the technical viewpoint we have to compute the
regular mode of the system where $H$ is a parameter with a given fixed value
and the role of the hamiltonian is played by $-M$.  Taking into account the
equation of motion (\ref{interiorequation}) we have now
\begin{equation} 
t' = 4M\log(\sqrt{\hat r}-\sqrt{2M})-4M\log(\sqrt{\hat r(0)}-\sqrt{2M})
\end{equation} 
while $p_c$ has still the form (\ref{pcsystem}).
Furthermore regularity is obtained by imposing that at $t'=0$,
$p_c=k$ with $k>0$.  Thus we have the further restriction 
\begin{equation} 0<k
= - \hat r(0)\log\frac{\sqrt{\hat r(0)}-\sqrt{2H}}{\sqrt{\hat r(0)}-\sqrt{2M}} 
\end{equation} 
where now $H$ is a constant parameter and $M$ is a function of $t'$.
The implicit equation for $M(t')$ we obtain is
\begin{equation}\label{Mequation}
k=(\sqrt{2M}+ T' (\sqrt{\hat r} -\sqrt{2M}))^2\log\frac{(\sqrt{\hat
r}-\sqrt{2M})T'}{(\sqrt{\hat r}-\sqrt{2M})T'-\delta_M}
\end{equation} 
with
\begin{equation}
T' = e^{-\frac{t'}{4 M(t')}};~~~~{\rm and}~~~~\delta_M=\sqrt{2H}-\sqrt{2M} 
\end{equation} 
For $t'\rightarrow \infty$ we have $M(t')\rightarrow H$
and thus expanding in the implicit variable $T'$ we have
\begin{equation}\label{deltaM}
\delta_M = c_M T'+O(T'^2)
\end{equation} 
where $c_M$ is easily computed from eq.(\ref{Mequation})
\begin{equation}
c_M =(1-e^{-\frac{k}{2H}})L';~~~~L'=\sqrt{\hat r}-\sqrt{2H}.
\end{equation} 
Repeating the same procedure as the one described in 
eq.(\ref{mainexpansion})
and squaring the relation (\ref{deltaM}) we obtain
\begin{equation}
M(t')= H -\sqrt{2H}c_M \tau' + \frac{t'}{2H}(c_M\tau')^2
\end{equation}
which is the analogue of eq.(\ref{Hoft}).
 Then the $v_k$ mode, i.e. the mode regular at the horizon is given by
\begin{equation}
e^{i(k\hat r(0)+\int_0^{t'} p_c \dot{\hat r}~dt'' + M(t') t')}.
\end{equation} 
Again we have
\begin{equation}
\frac{\partial (k\hat r(0)+\int_0^{t'} p_c \dot{\hat r}~dt'' + M(t') t')}
{\partial t'}= M(t')
\end{equation} 
so that the time dependence of the regular mode for fixed $\hat r$ is
\begin{equation}
e^{i(f(\hat r)+\int^{t'}M(t'')dt'')}.
\end{equation}
For $t'\rightarrow +\infty$ we have $M(t')\rightarrow H$. Subtracting such
background frequency we have the radiation mode regular at the horizon
\begin{equation}\label{Mregularradiatonmode}
e^{i(f(\hat r)+\int^{t'}M(t'')dt'' - H t')}.
\end{equation}
On the other hand we recall that the $u$-modes are characterized by well
defined values of $H$ and $M$.
On dealing with $M$ as hamiltonian one can use either the outer or the inner
gauge with no change in the results in $p_c$ as we have stressed in
Sect.(\ref{thereducedactionsec}). 
By subtracting again the background frequency $H t'$ from the total mode
\begin{equation}\label{Mumode}
e^{i(\int^{\hat r} p_c d\hat r +M t')}
\end{equation}
we obtain the $u$- radiation mode
\begin{equation}\label{Muradiatonmode}
e^{i(\int^{\hat r} p_c d\hat r -(H-M)t')}=e^{i(\int^{\hat r} p_c d\hat r
-\omega t')} 
\end{equation}
The physical identification of such $u$-mode with the mode (\ref{spacepart}) 
in the $t$ description is made by noticing that due to
the independence of $p_c$ of the scheme, the space part, i.e. the $\hat r$
dependence of the modes is the same with
\begin{equation}  
\omega\equiv H-M = \sqrt{\hat p^2(+\infty)+m^2}.  
\end{equation}  
Thus we have
for the time Fourier analysis at fixed $\hat r$ of the regular mode in terms 
of the $u$-modes
\begin{equation}  
\int dt' e^{-iH t'+ i\int^{t'} M(t'')dt''} e^{\pm i \omega t'}
\end{equation}
which is the same as eq.(\ref{exponentH}) with 
$-i\int^t H(t')dt' + i M t$ replaced by $-i H t'+ i\int^{t'} M(t'')dt''$.
Repeating the steps after eq.(\ref{exponentH}) we obtain for the saddle point
referring to the calculation of the $\beta$ coefficient the relation 
\begin{equation}  
H-M(t') = \sqrt{2H}\tau_1'-\frac{t'}{2H}(\tau_1')^2 = -\omega
\end{equation}
and for the saddle point value of the exponent
\begin{equation}\label{Mexponent}  
-i[4H\omega+t'\omega(1-\frac{\omega}{2H})].
\end{equation}
Again we need $t'$ to first order in $\omega$.
\begin{equation}\label{t1saddle}  
t'=-4H(1+\frac{\omega}{H})\log(-\frac{\omega}{\sqrt{2H}c_M});~~~~
{\rm Im} t'=-4\pi H (1+\frac{\omega}{H}).
\end{equation}
Combining eq.(\ref{Mexponent}) and (\ref{t1saddle}) we obtain the ratio
\begin{equation}\label{betaoveralphaM}
\left|\frac{\beta_{k\omega}}{\alpha_{k\omega}}\right|=e^{{-4\pi H\omega
(1+\frac{\omega}{2H})}}.   
\end{equation} 
 Now the asymptotic behavior of the time development of the regular mode in the
$t'$ representation is $e^{iHt'}$ and being $t'$ related to the interior mass
we have to identify the parameter $H$ as the mass of the remnant of the black
hole after the emission of the quantum of energy $\omega$ in agreement with
eq.(\ref{betaoveralphaH}) where as stated at the beginning of the present
section we have been working consistently to second order in $\omega/M$.

With regard to the validity of the late time expansion, it can be examined by
comparing the first order result, in eq.(\ref{saddlevalue}) with the second
order term. One easily finds that for the typical frequency of the Hawking
radiation i.e. $\omega/l_P^2$ of the order of magnitude $1/M$ with $l_P$
the Planck length, the ratio of the two terms is of order $l_P^2/M^2$ apart
logarithmic corrections thus holding for black holes of a few Planck masses or
higher.

\section{Conclusions}\label{conclusionssec}
  
In this paper we have examined a some issues related to the semiclassical
treatment of Hawking radiation. The main appeal of such an approach is the
possibility to take into account the back reaction i.e.  the fact that in the
transition the mass of the black hole varies. The description of the system by
means of a reduced hamiltonian is pivotal in this approach, and such a
reduction can be performed on rigorous ground at the classical level. The
introduction of the generating function $F$ of 
Sect.(\ref{thereducedactionsec}) simplifies greatly the
reduction process.  We proved that the canonical conjugate momentum $p_c$
which appears in the reduced action and which plays the major role in all
subsequent developments is invariant in the class of the Painlev\'e gauges. We
do not expect this to be true for all possible gauges but the Painlev\'e class
of gauges which are defined by the choice $L\equiv 1$ in the metric
(\ref{metric}), gives a good description of an asymptotic observer at space
infinity and has the good feature of being non singular at the horizon. There
is still a gauge freedom in the choice of $R$ but we proved that $p_c$ does
not depend on such a choice. The dynamics at the classical level con be
developed both using the exterior mass or the interior mass as hamiltonian and
these two descriptions are equivalent.

By interpreting the exponential of the reduced semiclassical action as modes of
the system as done in \cite{KW1} it is possible to perform the calculation of
the Bogoliubov coefficients. We have revisited in this context that late time
expansion which was introduced already in \cite{KW1}. We give a simplified
treatment of such an expansion leading to the back
reaction corrections both using the exterior or the interior mass as
hamiltonians. The late time expansion used in the treatment is expected to
hold for black holes mass of a few Planck masses or higher i.e. for massive
black holes. The gray body factor, as intrinsic in the semiclassical treatment
to lowest order, is just unity. In \cite{FM} the relation
eq.(\ref{Imofintegral}) was proven also for the emission of two interacting
shells (massive or massless) which during the evolution can cross. This
would point to the absence of correlations among emitted quanta \cite{parikh}
however up to now we have no mode interpretation of such a result.
  
\bigskip

\section*{Appendix A}\label{appendixsec}

In this appendix we derive the explicit expression for the conjugate canonical
momentum $p_c$ in the class of Painlev\`e gauges, showing its universality
within these gauges.
The general form of $p_c$ was given in \cite{FM}
\begin{equation}\label{pcoverR}
\frac{p_c}{R} = \Delta{\cal L}-\Delta{\cal B}=
\sqrt{\frac{2M}{R}}-\sqrt{\frac{2H}{R}}+\log\left[\frac{R'_+-W_+}{R'_--W_-}
\cdot \frac{1-\sqrt\frac{2M}{R}}{1-\sqrt\frac{2H}{R}}\right]
\end{equation}
where $R$ stays for $R(\hat r)$ which in our scheme equals $\hat r$, $R_+$ and
$R_-$ stay for the right and left derivatives of $R$ at $\hat r$.
We solve in terms of $R'_+$ using the two relations
(\ref{DeltaR1},\ref{DeltapiL}) 
\begin{equation}
R'_- = R'_++\frac{V}{R};~~~~W_- = W_++\frac{\hat p}{R}.
\end{equation}
By squaring the second equation we obtain
\begin{equation}\label{R1minusWminus}
R'_+\frac{V}{R}=A+\frac{\hat p}{R}\sqrt{{R'_+}^2 -1+\frac{2H}{R}}
\end{equation}
with
\begin{equation}
A=\frac{H-M}{R}-\frac{m^2}{2R^2}~.
\end{equation}
Squaring again we obtain for $\hat p$ the second oder equation
\begin{equation}
\left(1-\frac{2H}{R}\right)\frac{\hat p^2}{R^2}-2 A\sqrt{{R'_+}^2-1
+\frac{2H}{R}}\frac{\hat p}{R}-A^2 + m^2\left(\frac{R'_+}{R}\right)^2=0
\end{equation}
whose discriminant is given by
\begin{equation}
{R'_+}^2 C = {R'_+}^2 \left[\left(\frac{H-M}{R}\right)^2
+\frac{m^4}{4R^4}+\frac{m^2}{R^2}\left(\frac{H+M}{R}-1\right)\right].
\end{equation}
Then
\begin{equation}\label{explicitphat}
\frac{\hat p}{R}=\frac{AW_++R'_+\sqrt{C}}{1-\frac{2H}{R}}
\end{equation}
from which we see that $\hat p$ is a gauge dependent 
quantity i.e. a quantity which depends on $R'_+$. 
Using eq.(\ref{R1minusWminus}) with $\hat p$ given by the previous equation
substituting $R'_-$ and $W_-$ into eq.(\ref{pcoverR}) and multiplying 
both numerator and denominator by $R'_++W_+$ we obtain for $p_c$
\begin{eqnarray}
\frac{p_c}{R} =\sqrt{\frac{2M}{R}}-\sqrt{\frac{2H}{R}}+
\log \frac{\left(1-\sqrt\frac{2M}{R}\right) \left(1+\sqrt{\frac{2H}{R}}\right)}
{1-\frac{2H}{R}+\frac{H-M}{R}-\frac{m^2}{2R^2}-\sqrt{C}}=\\
=\sqrt{\frac{2M}{R}}-\sqrt{\frac{2H}{R}}+
\log \frac{1-\frac{2H}{R}+\frac{H-M}{R}-\frac{m^2}{2R^2}+\sqrt{C}}
{\left(1+\sqrt\frac{2M}{R}\right) \left(1-\sqrt{\frac{2H}{R}}\right)} 
\end{eqnarray}
showing the gauge independence of $p_c$ within the family of Painlev\`e
gauges.

For the discussion of the analytic properties of $p_c$ it is however simpler
to use the system eq.(\ref{pcouter},\ref{fundamentalH}) as done in \cite{FM}.

\vfill


\end{document}